\documentclass[a4paper,12pt]{revtex4}
\usepackage{amsmath}  
\usepackage{amssymb}  
\usepackage{dsfont}   

\begin{document}
\title{The quantum field theory interpretation of quantum mechanics}
\author{Alberto C. de la Torre}\email{delatorre@mdp.edu.ar}

\affiliation{Universidad Nacional de Mar del Plata}

\begin{abstract}
It is shown that adopting the \emph{Quantum Field} ---extended
entity in space-time build by dynamic appearance propagation and
annihilation of virtual particles--- as the primary ontology the
astonishing features of quantum mechanics can be rendered intuitive.
This interpretation of quantum mechanics follows from the formalism
of the most successful theory in physics: quantum field theory.
\end{abstract}
\maketitle
\section{Introduction}
After more than one century that Planck and Einstein made the first
quantum postulates\cite{plk,ein} and after 80 years that the
mathematical formalism of quantum mechanics was
established\cite{vneum}, the challenge posed by quantum mechanics is
still open. For many decades the situation was well described by R.
Feymnan when he said ``nobody understands quantum
mechanics''\cite{fey}. This is is perhaps no longer true due to the
achievements of the last decades. The lack of understanding was
compensated by the development of an extremely precise and esthetic
mathematical formalism; we did not know {\em what} quantum mechanics
is but we knew very well {\em how} it works. The development of the
very successful axiomatic formalism had the consequence that many
physicists where satisfied with the working of quantum mechanics and
did no longer tried to understand it. This attitude was favoured by
the establishment of an orthodox instrumentalist ``interpretation''
which, if we are allowed to put it in a somewhat oversimplified
manner, amounts to say ``thou shall not try to understand quantum
mechanics''.  Only a few authorities like Einstein, Schr\"odinger,
Planck, could dare not to accept the dogma and insisted in trying to
understand quantum mechanics\cite{jam}. Fortunately the situation
changed and the search for an interpretation of quantum mechanics
became an acceptable research subject. The roots for this change are
found in the pioneering work of Einstein Podolsky and
Rosen\cite{epr} which pointed out to some peculiar correlations in
the theory, followed by the work of Bell\cite{bell} that established
measurable consequences of them that were confirmed
experimentally\cite{exp}.

The increased activity in the field resulted in a very large number
of ``interpretations'' but, unfortunately, also in much confusion on
the precise meaning of producing an interpretation for quantum
mechanics. So, besides the Copenhagen or Complementarity
interpretation we can find Schr\"{o}dinger's field interpretation, the
de Broglie pilot wave interpretation, the hydrodynamic
interpretation, the many world interpretation, the modal
interpretation, the transactional interpretation, the coherent
histories interpretation, the Path Integral interpretation, the
causal (Bohmian) interpretation, the stochastic interpretation, the
statistical interpretation, the hidden variables interpretation, and
many other with ephemeral life. For more confusion, we should add to
the list the \emph{no interpretation} interpretations including many
instrumentalist claims that quantum mechanics does not need an
interpretation and just has to provide an algorithm for predicting
the results of experiments.

A study of the proposals shows that there is some confusion  about
what exactly is an interpretation. Unfortunately, it seems that any
new idea about some general feature of the theory, or some
metaphorical model, or an alternative mathematical formalism, is
called ``an interpretation''. This situation may result in a sterile
proliferation of interpretations. In order to limit this growth and
to clarify this issue, we can choose set of quite reasonable minimal
requirements that a proposal should fulfill in order to be called an
interpretation.
\begin{description}
    \item[Realism] {\em Every interpretation of quantum mechanics
must be realist.} This amounts to the philosophical postulate of the
objective existence of reality independent of any observation,
although any act of observation may produce strong, or even
unpredictable, effects.

The objects of study of quantum mechanic, the \emph{Quantum
Systems}, is an abstraction of reality defined by a set of
observables that we use to build models of reality. The knowledge
that physics has provided about these models of reality forces us to
accept that the quantum system may have properties by far more
sophisticated than the ones detected by our sense perception, and
that their behaviour may contradict our classical intuition. In
other words, physics has shown that {\em naive realism} is wrong.

Any interpretation must be realist because in an interpretation we
associate the results obtained from the theory or from the
experiments with some existent objects. An interpretation of a
theory becomes meaningless without the existence of the objects to
which it is applied. Many physicist may be surprised by the
necessity to state such a postulate because they may take it as
obvious. However it is convenient to state it explicitly because
there are ideologies and epistemological schools that question the
postulate of realism.
\par
The search for an interpretation of quantum mechanics implies the
acceptance  of another philosophical postulate: {\em Nature not only
exists but it can also be known, at least in an ever increasing
approximation, by means of physical theories.} That is, quantum
mechanics can give us information about reality, even if it is
affected by inherent uncertainties or indeterminacies. Therefore,
quantum mechanics is telling us something about nature and not
merely about the observations that we make of nature.
    \item[Physical Space-Time]
{\em Every existent physical system is embedded in space-time} and
is associated to a domain of it according to the equations of motion
of the theory. This space has some geometrical structure allowing
the assignment of coordinates and there are mathematical
transformations of coordinates relating different frames of
reference. These transformations may depend on several physical
constants in a way that, under the appropriated limit, Poincar\'{e},
Lorentz or Galilei transformations are obtained.
\par
It is important to notice that this requirement {\em does not} say
that physical space is a four dimensional Minkowski space or a
Riemann space with curvature or a three dimensional Euclidian space
and a one dimensional time. The dimension and geometrical structure
of physical space-time  may be anything, provided that in the
appropriated limit the spaces of classical physics or of special or
general relativity are reached. Considering the difficulties
encountered by all attempts to find an interpretation of quantum
mechanics in the usual Minkowski or Euclidian spaces, we may expect
that, perhaps, the advent of a definite interpretation for quantum
mechanics will require a radical proposal of some unexpected
geometry for physical space-time. For this reason, it is important
that this requirement should not restrict the possible geometrical
structures that may be necessary to assign to physical space.
\item[Primary Ontology] {\em Every interpretation of quantum mechanics must propose a
primary ontology}. A possible reason for the difficulties in finding
an interpretation of quantum mechanic was perhaps a wrong choice of
an ontology from the beginning. In many attempts, either a
\emph{particle} or a \emph{field} ontology was assumed. These two
choices are very successful in classical physics but clearly fail
with quantum mechanics. In order to overcome these failures, the
concepts of particle-wave duality was introduced as a manifestation
of the more general principle of complementarity.

The requirement of a primary ontology means that the interpretation
must clearly state what are the basic existent things in physical
space, that are the carriers of energy and momentum or of other
observable properties. In early interpretations, these primary
ontology where particles or fields and these entities were endowed
with non classical properties like the complementary presence of
dual properties. Whatever this primary ontology is, it must exist in
physical space as carrier of energy-momentum. This requirement
excludes the ``histories'' or the ``correlations'' from being the
primary ontology. There have been several attempts of
interpretations based on different choices for the primary ontology.
Without many details we just mention some of them\cite{int}. The
well known probability interpretation of the wave function $\psi(x)$
proposed by Max Born favours a particle ontology. In this case
$\psi(x)$, a Hilbert space element, does not carry energy and is not
really existent in physical space. Opposite to it, we can find
Schr\"odinger interpretation proposing that only the wave function
has objective existence. A hybrid interpretation was proposed by L.
de Broglie with his ``double solution'' suggesting a mixed particle
{\em and} field ontology. Another idea originated by L. de Broglie
is based on a particle ontology with a pilot wave determining its
motion. This interpretation was successfully taken by D. Bohm in his
causal quantum mechanics (Bohmian mechanics).
\end{description}

These criteria are not satisfied by several proposals mentioned
above. In this work we will see that an interpretation of quantum
mechanics based on an entity different from particles or fields that
we name \emph{Quantum Field} can be adopted providing a somewhat
intuitive understanding. We will try to show that most astonishing
features of quantum mechanics can be explained as a natural
consequence of the ontology suggested by quantum field theory based
on a permanent creations and annihilation of virtual particles and
antiparticles. Indeterminacies, nonlocality consequences of
superposition, individuality entanglement of identical particles,
and many other features of the quantum system, not understood in the
particle or in the field ontology, become natural features of the
quantum field built by virtual particles.
\section{The ontology of the quantum field}
There exist a set of physical entities, called \emph{Elementary
Particles}, characterized by different values of some observable
properties.  They are listed in the \emph{Standard Model} and are
identified as electrons, neutrinos, quarks, photons, etc. Associated
with each elementary particle we define a physical system called
\emph{Virtual Particle} consisting in the creation of the particle
at some space-time point, its propagation with definite
energy-momentum and its annihilation at another space-time point.
Opposed to virtual reality in computer simulations, virtual
particles do exist in reality but with ephemeral live. These virtual
particles exist and have observed empirical consequences as in the
Casimir effect or in the Lamb shift. They can not be permanent
because they do not satisfy the energy momentum relation
$m^{2}=E^{2}-P^{2}$ (they are off the mass shell) and they can
propagate in space-like trajectories. This fact has the astonishing
consequence of the necessary existence of \emph{antiparticles}: for
a virtual particle propagating in a space-like trajectory between
the times $t_{1}$ and $t_{2}$ $(t_{1}<t_{2})$ in a reference system
$S$ there is a Lorentz transformation to $S'$ where the
corresponding \emph{antiparticle} is propagating between $t'_{2}$
and $t'_{1}$ $(t'_{2}<t'_{1})$.

The \emph{Quantum Field} is a physical entity extended and evolving
in space-time according to specific equations of motion
(Schr\"{o}dinger, Dirac, Klein-Gordon) made by an infinite set of
virtual particles. At every space-time point the amplitude of the
intensity of the field denotes the existence of particles,  and
likewise, the field provides the amplitude for realization of every
energy-momentum value. The quantum field is the primary ontology
with permanent existence; however it is not simple and elementary
because it is composed by the superposition of virtual particles. In
this view, Feynman graphs represent not only a term in a
perturbation expansion but they describe real processes occurring in
physical space. All these features are compatible with the
mathematical formalism that will be described in the following
section.
\section{Minimal Quantum Field Theory}
In this section a minimal version of quantum field theory is
presented containing only those features required for the
understanding of quantum mechanics based on its ontology. For this
purpose we don't need to consider specific spin values of the
particles described by the theory neither do we need to describe the
details of the interactions between different particles involving
advanced mathematical techniques. Therefore we consider only the
main features of quantum fields and we avoid the mathematical
complications that sometimes blur the essential features of the
theory. There are excellent books where quantum field theory is
presented in all rigour and details\cite{wein}.

The physical system that the quantum field describes is an
indefinite number of some type of particle (electron, quark, photon,
etc.) in its space-time evolution. The \emph{state} of such a
system, that is, that mathematical entity allowing us make any
prediction concerning the observables, is an element of a Fock space
$\mathcal{H}$ built as the orthogonal sum of Hilbert spaces
\begin{equation}\label{fock}
 \mathcal{H}=\mathcal{H}^{0}\oplus\mathcal{H}^{1}\oplus\mathcal{H}^{2}\oplus\ldots\oplus\mathcal{H}^{n}\oplus\ldots
\end{equation}
where
\begin{equation}\label{Hilbn}
\mathcal{H}^{n}=\mathcal{H}\otimes\mathcal{H}\otimes\ldots
\end{equation}
is the Hilbert space for an $n =1,2,\ldots$ identical particle
system and $\mathcal{H}^{0}$ contains only one element: the
(normalized) vacuum state $\psi_{0}$ (not to be confused with the
null element of any Hilbert space).

A useful basis in Fock space is given by the eigenvectors of the
position operator of the particles $\{\varphi_{x_{1},x_{2},\ldots
x_{n}}\ \forall n \}$ built as linear combinations of all label
permutations of the element $\varphi_{x_{1}}\otimes\varphi_{x_{2}}
\otimes\ldots\otimes\varphi_{x_{n}}$ (in $\mathcal{H}^{n}$) such
that the resulting state is totaly symmetric or anti-symmetric when
the particles described are bosons or fermions respectively.

An interesting feature of quantum field theory, that is absent in
non relativistic quantum mechanics, is the possibility of states
with an indefinite number of particles, described by a superposition
of Hilbert space elements belonging to different subspaces of
Eq.(\ref{fock}). An important example of this, appears in the
quantum field for photons: a state with an \emph{exact} number of
photons has \emph{zero} expectation value for the electric and
magnetic field observables and only with states that are \emph{not}
eigenvector of the number operator can we observe nonzero values of
the electric and magnetic fields. Other interesting states, also
with non definite number of particles, are the \emph{coherent
states} (eigenvectors of the annihilation operator) that turn out to
be the states closest to the classical behaviour of the system.

A central feature of quantum field theory is the description of
spontaneous creation and annihilation of particles by means of
operators that connect the Hilbert spaces of Eq.(\ref{fock})
increasing or decreasing the number of particles. More precisely,
consider some one particle state $\varphi\in\mathcal{H}^{1}$
corresponding to some property of the particle, that is, $\varphi$
is an eigenvector of some observable. We define now a \emph{creation
operator} $A^{\dag}$ such that $A^{\dag}\psi_{0}=\varphi$ and when
applied to any state of $\mathcal{H}^{n}$ results in an element of
$\mathcal{H}^{n+1}$ (properly symmetrized or anti-symmetrized) with
an extra particle in the state $\varphi$. Correspondingly, $A$ is
the annihilation operator for a particle in the state $\varphi$.

If we consider now a set of creation operators
$\{A^{\dag}_{\alpha}\}$ corresponding to a basis
$\{\varphi_{\alpha}\}$ in $\mathcal{H}^{1}$, then we can obtain any
multiparticle state in Fock space by the application of these
operators to the vacuum state. Furthermore, not only the states, but
also all operators in Fock space can be expressed in terms of
creation and annihilation operators making them ubiquitous in the
formalism of quantum field theory. For instance, the operator
$A^{\dag}A$ is related with the number of particles in the state
$\varphi$ (zero or one for fermions) and therefore
$\sum_{\alpha}A^{\dag}_{\alpha}A_{\alpha}$ is the operator for the
total number of particles in the system.

The symmetrization requirements of the states imply that the
creation and annihilation operators must satisfy commutation (for
bosons) or anti-commutation (for fermions) relations:
\begin{equation}\label{CommRel}
 [A_{\alpha},A^{\dag}_{\beta}]_{\pm}=\delta_{\alpha,\beta}\mathds{1}
 \ ,
 [A^{\dag}_{\alpha},A^{\dag}_{\beta}]_{\pm}=0 \ ,
 [A_{\alpha},A_{\beta}]_{\pm}=0 \ .
\end{equation}

As said before, one of the great achievements of quantum field
theory was the prediction of the existence of antiparticles.
Furthermore we will see that their existence is necessary in order
to satisfy relativistic causality. Therefore, in the formalism, it
is necessary to include operators $\bar{A}^{\dag}$ and $\bar{A}$ for
creation and annihilation of antiparticles. Since antiparticles and
particles annihilate each other (except when they are identical) the
total number of particles in a given state corresponds to the
operator $A^{\dag}A-\bar{A}^{\dag}\bar{A}$. The question naturally
arises whether the creation of an antiparticle is equivalent to the
annihilation of a particle, that is, whether $\bar{A}^{\dag}=A$ and
$\bar{A}=A^{\dag}$. If this were so, the total number of particles
in a given state would be associated with the operator
$A^{\dag}A-AA^{\dag}$, but this is always $-1$ for bosons: an absurd
result. Therefore for boson fields $\bar{A}^{\dag}\neq A$ and
$\bar{A}\neq A^{\dag}$ and we must express the quantum field using
both set of operators whereas for fermion fields we may do it with
just one type of creation and annihilation operators. There is
however an exception in this argument: the case where the bosons are
neutral (with respect to electric and all other charges) and
identical to the antiparticles (photons, for instance). In this case
we can make indeed $\bar{A}^{\dag}=A$ and $\bar{A}=A^{\dag}$.

Consider now the creation and annihilation operators
$B_{\beta}^{\dag}$ and $B_{\beta}$ related with a basis
$\{\phi_{\beta}\}$ in $\mathcal{H}^{1}$ different from the basis
$\{\varphi_{\alpha}\}$ created by $\{A_{\alpha}^{\dag}\}$. Using the
unitary transformation among the bases we readily obtain a relation
among creation and annihilations operators:
\begin{equation}\label{BArel}
 B_{\beta}^{\dag}=\sum_{\alpha} \langle\varphi_{\alpha},
 \phi_{\beta}\rangle A_{\alpha}^{\dag}\quad\ ,\quad
B_{\beta}=\sum_{\alpha} \langle\phi_{\beta},\varphi_{\alpha}
 \rangle A_{\alpha}\ ,
\end{equation}
and same equations for antiparticle creation and annihilation.

We are now ready for the presentation of the main tool of quantum
field theory: this is, essentially, the equations above but relating
the creation and annihilation operators for \emph{position}
eigenstates with those for \emph{momentum} eigenstates. Let then
\begin{equation}\label{Annfield}
\mathbf{\Psi}(x)=\sum_{p}\left(u(x,p)A(p)+v(x,p)\bar{A}^{\dag}(p)\right)
\end{equation}
be the annihilation  operator for a particle in the space-time
location $x=(t,\mathbf{x})$ given in terms of the annihilation of
particles (and creation of antiparticle) with all possible energy
momentum $p=(E,\mathbf{p})$. The corresponding creation operator is
obtained by hermitian conjugation. This general expression is
schematic and several comments are due to make it clear.
\begin{enumerate}
 \item The variables $x$ and $p$, playing the role of the
indices $\beta$ and $\alpha$, are  continuous and therefore the
summation symbol must be understood as an integral with a Lorentz
invariant integration measure. Furthermore, this summation should
also involve the spin degree of freedom that we have suppressed in
this schematic treatment.
\item The operator $\mathbf{\Psi}(x)$ and the complex functions $u(x,p)$ and
$v(x,p)$ have implicit several components in the different cases:
one for scalar (spin zero) particles, three for vector (spin one
massive) particles, four for Dirac spinors, sixteen for tensors,
etc. and have the appropriate behaviour under Lorentz
transformations.
\item In all the cases mentioned above, the operator $\mathbf{\Psi}(x)$ satisfy
some equation of motion for the field (Klein-Gordon, Dirac, etc.).
There are in fact two approaches in the presentation of quantum
field theory: in one, as suggested here, we start with particles and
obtain the equation of motion of the operator fields and in the
other approach we start from a Lagrangian and find the solutions of
the Euler-Lagrange equations to represent the particles.
\item In the cases where particles and antiparticles are identical
we can replace $\bar{A}^{\dag}(p)=A(p)$.
\item As mentioned, this expression is schematic and the exact form,
suitable for calculations, can be found in appropriate
books\cite{equ}.
\item The commutation or anti-commutation relations
Eq.(\ref{CommRel})  for the fields (adapted for continuous
variables) vanish when evaluated at points $x$ and $y$ such that
$x-y$ is space-like. This important requirement of relativistic
causality could not be satisfied without antiparticles.
\end{enumerate}

The interactions among particles is introduced in quantum field
theory by means of gauge fields with creation and annihilation of
the carriers of the interactions: photons, weak vector bosons,
gluons and gravitons. When possible, Feynman diagrams represent all
perturbation orders of the interaction involving creation,
propagation and annihilations of gauge bosons and particles.

As suggested above, the formalism of quantum field theory favours
the interpretation based on a permanent creation and annihilation of
particles and antiparticles at every location with a given
intensity. In order to see how the formalism supports this
interpretation let us consider the description that quantum field
theory makes of some very simple physical systems. Let
$\psi\in\mathcal{H}^{1}$ be the state of a one particle system at
some time. If we expand it in the basis $\{\varphi_{x}\}$
corresponding to the eigenvectors of the position operator, we have
$\psi=\sum_{x}f(x)\varphi_{x}$. Now we write $\varphi_{x}$ given by
the creation field applied to the vacuum.
\begin{equation}\label{stateonepart}
 \psi=\sum_{x}f(x)\mathbf{\Psi}^{\dag}(x)\ \psi_{0}\ .
\end{equation}
This suggests the interpretation that $f(x)$ denotes the
\emph{intensity} of the quantum field of the one particle system.
That is, \emph{at any location $x$, a particle is created from the
vacuum with an intensity $f(x)$.} In order to support this, let us
calculate the number of particles at the location $x$ for this
state, that is, the expectation value of
$\mathbf{\Psi}^{\dag}(x)\mathbf{\Psi}(x)$:
\begin{eqnarray}
\nonumber
 \left\langle\psi\ ,\ \mathbf{\Psi}^{\dag}(x)\mathbf{\Psi}(x)\psi
\right\rangle &=&
\left\langle\sum_{x'}f(x')\mathbf{\Psi}^{\dag}(x')\ \psi_{0} \ ,\
\mathbf{\Psi}^{\dag}(x)\mathbf{\Psi}(x)
\sum_{x''}f(x'')\mathbf{\Psi}^{\dag}(x'')\ \psi_{0} \right\rangle\\
 &=& \sum_{x'}\sum_{x''}f^{\ast}(x')f(x'')\left\langle \psi_{0}\ ,\
\mathbf{\Psi}(x') \mathbf{\Psi}^{\dag}(x)\mathbf{\Psi}(x)
\mathbf{\Psi}^{\dag}(x'')\ \psi_{0} \right\rangle\ .
\end{eqnarray}
Now, using the commutation or anti-commutation relations we can
shift all the creation field operators to the left (that is,
expressed in ``normal order'') and considering that the annihilation
operator applied to the vacuum produces the null element, we obtain
\begin{equation}\label{numberofpartinx1}
\langle\psi\ ,\ \mathbf{\Psi}^{\dag}(x)\mathbf{\Psi}(x)\psi \rangle
= \sum_{x'}\sum_{x''}f^{\ast}(x')f(x'')
\delta_{x,x'}\delta_{x,x''}\langle \psi_{0}, \psi_{0} \rangle
=|f(x)|^{2}\ .
\end{equation}

Let us consider a virtual particle created at the location $x$ with
an intensity given by the complex function $f(x)$. The modulus
squared of this function gives then the existential weight
(probability) for the particle at $x$. However, this function must
also contain information indicating that the virtual particle
belongs to a collective of virtual particles that make up the field
for the real particle: it must contain information about all other
observables. This information is contained in a holistic way
involving all values of $x$. For instance, the intensity for the
creation of a virtual particle with momentum $p$ is given by
$g(p)=\sum_{x}f(x)\langle\phi_{p},\varphi_{x}\rangle$ where
$\langle\phi_{p},\varphi_{x}\rangle$ is the internal product between
the eigenvectors of position and momentum.

As a generalization of Eq.(\ref{stateonepart}) we have the most
general state in Fock space
\begin{equation}\label{GeneralState}
 \psi=\sum_{n}
 \sum_{x_{1},x_{2},\ldots, x_{n}}
 F_{n}(x_{1},x_{2},\ldots, x_{n})
 \mathbf{\Psi}^{\dag}(x_{1})
 \mathbf{\Psi}^{\dag}(x_{2})\ldots
 \mathbf{\Psi}^{\dag}(x_{n})
 \ \psi_{0}\ .
\end{equation}
Any physically relevant quantity (transition amplitude, scattering
matrix, etc.) can be given in terms of internal products among two
states like the one above. That is, it will involve the vacuum
expectation value of products like $\mathbf{\Psi}(x_{1})
 \mathbf{\Psi}(x_{2})\ldots
 \mathbf{\Psi}(x_{n})\mathbf{\Psi}^{\dag}(x_{n+1})
 \mathbf{\Psi}^{\dag}(x_{n+2})\ldots
 \mathbf{\Psi}^{\dag}(x_{n+m}) $ for all $n,m,$ and $x_{i}$. In the
 formalism of quantum field theory, every physically relevant
 quantity or process is expressed in terms of creation and annihilation of
 virtual particles and in the proposed ontology this restless
 activity is assumed to occur in reality.

The commutation and anti-commutation relations of Eq.(\ref{CommRel})
were motivated by the symmetrization requirements of identical
particles states. However, the first of these relations allows an
interesting interpretation in agreement with the proposed ontology
for quantum field theory: for any location $x$, the identity
operator $\mathds{1}$ can be written as
$\mathds{1}=\mathbf{\Psi}(x)\mathbf{\Psi}^{\dag}(x)\ \pm
\mathbf{\Psi}^{\dag}(x)\mathbf{\Psi}(x)$, that is, as a combination
of creation and annihilation of particles. Applied to any state
(including the vacuum),
$\psi=\mathbf{\Psi}(x)\mathbf{\Psi}^{\dag}(x)\psi\pm
\mathbf{\Psi}^{\dag}(x)\mathbf{\Psi}(x)\psi $, suggesting that any
state can be thought as resulting from a permanent creation and
annihilation of particles.
\section{individuality loss}
One of the fundamental features of reality discovered by quantum
mechanics is the \emph{individuality loss}. In our perception of
macroscopic objects we take for granted that their individuality is
conserved: if we look at a stone, close our eyer for a second, and
observe it again, we never doubt that we are dealing with \emph{the
same} stone. This anthropocentric conviction can not be extrapolated
to the microscopic world. Identical \emph{classical} systems have an
individuality that is conserved through the time evolution and
interaction with other system (this conservation of individuality
corresponds to the concept of \emph{conatus} in antique Greek
philosophy). So classical systems, even when they are ``identical'',
can be assigned an individual identity that is conserved: they can
have a name, an ID number, a licence plate. Quantum mechanics
requires a drastic conceptual change: \emph{the individuality loss}.
A set of five identical ``classical'' atoms is countable (five in
total) and numerable (the atom number one, the number two,\ldots)
but real atoms, necessarily described by quantum mechanics, are
countable but not numerable. The individuality of the particles is
entangled with the individuality of all other identical ones in the
universe (although ``for all practical purposes'' a cluster
decomposition isolating a particular system from the rest is
possible to an extremely good approximation\cite{dlTM}).

Consider, for instance, two different states $\xi$ and $\eta$
belonging to the Hilbert space for one particle system
$\mathcal{H}^{1}$. The state of a two identical particle system
belongs to $\mathcal{H}^{2}=\mathcal{H}^{1}\otimes\mathcal{H}^{1}$
and the individuality entanglement requires a state proportional to
$\ \xi\otimes\eta\ \pm\ \eta\otimes\xi$ symmetric (for bosons) or
antisymmetric (for fermions). Notice the formal similarity of this
state with EPR-Bell entangled states where two subsystems exhibit
correlations that have been extensively studied. There are cases,
however, where the subsystems are not entangled and a separated
treatment is possible. On the contrary, the \emph{individuality}
entanglement in identical particle states is a distinctive feature
of quantum mechanics that can not be avoided.

It turns out that individuality entanglement is not just an
interesting feature but is one of the essential features of quantum
physics and therefore any complete interpretation of quantum
mechanics must provide a rational explanation or understanding for
the individuality entanglement. In the ontology suggested by quantum
field theory the individuality loss is very natural because in this
interpretation we are not dealing with one, or two, or many
particles as individual entities. For instance, the field for a one
electron system, or for several electrons system, is made up by the
permanent creation propagation and annihilation of virtual particles
that are not assigned to any of the individual electrons of the
system: in a two electron field there is no way to differentiate one
electron from the other because they are both simultaneously made by
an active background of ephemeral virtual particles with a mean
value of \emph{two} for the particle number observable, but each
virtual component  of the field is not assigned to any one the
electrons.
\section{Distributions in nonrelativistic quantum mechanics}
The predictions of non relativistic quantum mechanics are presented
in the form of distributions for the eigenvalues of the operator
associated with an observable. That is, for a system in a state
$\psi$, the theory provides for any observable $L$ with eigenvectors
$\{\varphi_{\lambda}\}$ (associated with the eigenvalue $\lambda$)
the distribution
$\rho(\lambda)=|\langle\varphi_{\lambda},\psi\rangle|^{2}$ that can
be tested empirically. Unfortunately, the name ``probability
distribution'' is irreversibly installed in quantum mechanics for
this function, although this is a misnomer because this quantity
does not satisfy all the requirements that the mathematical theory
requires for a probability. There are historical reasons for this
name in addition to the fact that it is measured experimentally as
if it were a probability, that is, by the frequency of appearance of
each eigenvalue. Anyway, other names for it have been proposed like
\emph{``pseudo-probability''} or, more recently, \emph{``existential
weight''}\cite{dlt1} but with little hope for acceptance.

One question that has dominated the research in the foundations of
quantum mechanics is the nature of this distribution. There are
basically two options: an ontological or a gnoseological
interpretation. We say that the existential weight has a
\emph{gnoseological} interpretation if we assume that the system in
its reality posses some definite value for the observable ---the
putative value--- but we are unable to know it because the theory is
unable to predict it: the system has a definite value but we can not
know it. The indeterminacy resides in our knowledge of the reality
of the system that has some hidden features. In this interpretation
the question immediately arises about the existence of a better
theory that can predict the exact value, the so called hidden
variable theories. In the opposite interpretation, the
\emph{ontological}, we accept that the observables are diffuse by
nature and do not assume precise values: quantum mechanics is a
complete theory and the indeterminacies are \emph{in} the reality of
the system and not in our knowledge if it.

At first sight, the gnoseological interpretation appears to be less
traumatic and was intensively investigated after the appearance of
the crucial paper of Einstein, Podolsky and Rosen\cite{epr}.
However, theoretical and empirical developments put severe
restrictions in the theories with hidden reality and many experts
today favour the ontological interpretation of the indeterminacies.
In fact, the Bell\cite{bell1} and Kochen-Specker\cite{koch} theorem
show that the existence of non contextual putative values for
commuting observables enters in contradiction with the
\emph{formalism} of quantum mechanics. Much more definitive, the
experimental violation\cite{exp} of Bell inequalities\cite{bell}
show that the existence of such non contextual putative values is in
contradiction with \emph{reality}. Context independence means that
the putative value of an observable does not depend on what other
commuting observables are being considered; a very reasonable
assumption because the context can be decided by theoretician at his
office and this should not change the reality of a physical system.

In the quantum field theory interpretation of quantum mechanics the
indeterminacies are ontological: the quantum field of a particle is
extended in space with an existential weight for the location of the
particle at any position given by the amplitude of the intensity for
the creation of particles at that point. Similarly every momentum
value is realized with an existential weight given by the
corresponding intensity of the field. Position and momentum of the
system described by the quantum field are diffuse and are related by
Fourier transformation that is a realization of a symmetry arising
from the equivalence of the description of the system by means of
its location or its movement (being and becoming
symmetry)\cite{dlt9}.
\section{position-momentum correlations}
The interpretation of the quantum field as permanent creation and
annihilation of virtual particles provide a very intuitive view of
the position-momentum correlations of a particle\cite{dlt8}. In
order to see this, let us consider the simplest system consisting of
one free particle moving in one dimension.  The position-momentum
correlation is defined as
\begin{equation}\label{Corr}
    C=\frac{1}{2}(XP+PX)\
\end{equation}
with commutation relations
\begin{equation}\label{ComRel}
   [X,C]= i\hbar X\ \hbox{ and }\ [P,C]= -i\hbar P \ .
\end{equation}

Let us imagine the virtual components of the quantum field created
at a location at ``the right'' of the one dimensional distribution
for position  $\rho(x) $, that is, with a \emph{positive} value for
the observable $X-\langle X\rangle$. If these components are moving
with momentum smaller than the mean value, that is, with
\emph{negative} value for $P-\langle P\rangle$ the relative motion
will be towards the center of the field and the distribution will
shrink. Similarly, the components created at the left and moving to
the right have the two offsets $X-\langle X\rangle$ and $P-\langle
P\rangle$ with different sign, that is, their (symmetrized) product
is negative.

For simplicity, let us assume that in this state we have $ \langle
X\rangle=\langle P\rangle =0$ (the general state is obtained with
the translation and impulsion operator). Therefore, the product of
the two offsets in position and momentum is precisely the
correlation observable and the previous argument means that if the
correlation is negative then the space distribution shrinks. We can
prove this with rigour: let us calculate the time derivative of the
width of the distribution $\Delta^{2} x = \langle X^{2}\rangle $. In
the Heisenberg picture, assuming a nonrelativistic hamiltonian
$H=P^{2}/2m$, we have
\begin{equation}\label{shrink}
 \frac{dX^{2}}{dt}=\frac{-i}{\hbar}[X^{2},H]=\frac{-i}{2\hbar m}[X^{2},P^{2}]
 =\frac{1}{m}(XP+PX)=\frac{2}{ m}C.
\end{equation}
Taking expectation values we conclude that states with negative
correlation shrink and states with positive correlation expand, as
expected from the heuristic argument given above based on the
reality of the virtual components of the field.

The momentum distribution for a free particle is time independent
and if the field is shrinking, that is, with negative correlation,
we are approaching the limit imposed by Heisenberg indeterminacy
principle. This principle will not be violated because the
correlation will not remain always negative: at some time it will
become positive and the state will begin to expand. In fact, we can
prove that the correlation is never decreasing in time:
\begin{equation}\label{corrtimeincr}
 \frac{dC}{dt}=\frac{-i}{\hbar}[C,H]=\frac{-i}{4\hbar m}[XP+PX,P^{2}]
 =\frac{1}{m}P^{2}=2H,
\end{equation}
and this is a nonnegative operator. If the field is shrinking, at
some later time it will be spreading. Gaussian states of this sort
have been reported\cite{rob} in a very comprehensive paper.
\section{superposition}
The principle of superposition establishes that if $\psi_{1}$ and
$\psi_{2}$ are two possible states of a system then
$\psi\propto\psi_{1}+\psi_{2}$ is another possible state. This
principle is a necessary consequence of the linear structure of the
Hilbert space of states and of the linearity of the causal evolution
of the system that preserves the superpositions. Another way of
looking at it, is to think that any state $\psi$ can be decomposed
in an infinite number of ways into components involving all Hilbert
space element not orthogonal to the given state, that is, related to
all properties not incompatible with the one fixing the state. In
this way, the state contains information about all possible
properties of the system. A useful application of the principle of
superposition corresponds to the mathematical possibility of
expanding any state in a basis. Physically, this expansion provides
the content of the state concerning all eigenvalues of an
observable. Notice however that \emph{states} are superposed but not
the \emph{properties} of the system associated with them. In fact,
if $\psi_{1}$ and $\psi_{2}$ are eigenvectors of some observables
corresponding to two different properties of the system, then $\psi$
is \emph{not} an eigenvector corresponding to any one of these
properties.

Let us consider the quantum field of a particle in a state
$\psi_{1}$. According to the ontology proposed, this field is build
by a permanent creation and annihilation of virtual particles. The
same can be said for the state $\psi_{2}$. Let us assume now that
these two states correspond to quantum fields separated in physical
space. In this case, the superposition
$\psi\propto\psi_{1}+\psi_{2}$ corresponds to a quantum field where
virtual particles are created somewhere and annihilated far away
providing some sort of ``quantum rigidity'' or non-separability that
played a relevant role in the EPR argument.

An other interesting consequence of the superposition of states of
compound systems is entanglement that will be mentioned next.
\section{entanglement}
Entanglement is one of the most remarkable features of
nonrelativistic quantum mechanics exhibiting strong correlations
between unrelated observables in compound systems. Most physical
systems are compound, in the sense that they can be decomposed in
subsystems, sometimes corresponding to separate physical systems
(like an electron and a proton in a hydrogen atom) or to different
degrees of freedom of one system (like spin and location of the same
particle, or different space coordinates).

The Hilbert space for the states of the compound system
$S=(S_{A},S_{B})$ is a tensor product structure
$\mathcal{H}=\mathcal{H}_{A}\otimes\mathcal{H}_{B}$. Consider two
different properties of the subsystem $S_{A}$ (for instance, spin
1/2 in two different orientations) denoted by $A_{1}$ and $A_{2}$
corresponding to the states $\varphi_{1}$ and $\varphi_{2}$ that
may, or not, be orthogonal.  Consider also another unrelated pair of
properties $B_{1}$ and $B_{2}$ of $S_{B}$ associated with $\phi_{1}$
and $\phi_{2}$ (for instance, located here or there). Furthermore,
imagine two possible states of the system:
$\varphi_{1}\otimes\phi_{1}$, corresponding to the simultaneous
appearance of the properties $A_{1}$ and $B_{1}$ and the other
state, $\varphi_{2}\otimes\phi_{2}$, corresponding to the appearance
of the properties $A_{2}$ and $B_{2}$. The superposition,
$\varphi_{1}\otimes\phi_{1}+\varphi_{2}\otimes\phi_{2}$, is an
entangled state of the system. In this state, none of the properties
$A_{1},A_{2},B_{1},B_{2}$ are objective (in the sense that the state
is \emph{not} an eigenvector corresponding to any of these
eigenvalues) but there are strong quantum correlations among them
because the observation of one property, say $A_{1}$, forces the
appearance of $B_{1}$ although they may be totaly unrelated (like
spin and location). In entangled states all sort of astonishing
quantum effects appear, like violations of Bell's inequalities,
Einstein-Podolsky-Rosen (so called) paradox, Schr\"{o}dinger cat,
nonlocality, teleportation, quantum cryptography and computation,
etc. The principle of superposition, that generates the
entanglement, contains perhaps the central essence of
nonrelativistic quantum mechanics and almost all pondering
concerning its foundations involve entangled states.
\section{MEASUREMENT}
The understanding of the measurement process in quantum mechanics is
very controversial but can be described following the scheme devised
by von Neumann\cite{vneum} and the London Bauer theory\cite{LonBau}
without intervention of the observer conscience and with the
physical process of decoherence\cite{Decoh} replacing the
unnecessary ``collapse''.

In order to describe the measurement process let us consider a
physical system $S$ in a state expanded in the eigenvectors
$\varphi_{\lambda}$ of an observable $L$ to be measured:
$\psi=\sum_{\lambda}f(\lambda)\varphi_{\lambda}$. The measurement
apparatus is another quantum system $S_{A}$ that can be in a set of
states $\{\phi_{\lambda}\}$ corresponding to the reading $\lambda$
in its display. During the measurement, both system interact and the
compound system  $(S,S_{A})$ is set in an entangled state
$\sum_{\lambda}f(\lambda)\,\varphi_{\lambda}\otimes\phi_{\lambda}$.
Although the apparatus is treated as a quantum system, it is
macroscopic, has a large energy $E_{A}$ and could be treated
classically. This means that after the interaction, in an extremely
short decoherence time that can be estimated as
$\frac{\hbar}{E_{A}}$, the system makes a transition from the pure
state to a mixed state with \emph{classical} probabilities:
\begin{equation}\label{decoherence}
 \sum_{\lambda}f(\lambda)\,\varphi_{\lambda}\otimes\phi_{\lambda}
 \longrightarrow
 \sum_{\lambda}|f(\lambda)|^{2}P_{\lambda}\ ,
\end{equation}
where $P_{\lambda}$ is a projector in the state
$\varphi_{\lambda}\otimes\phi_{\lambda}$.

In the decoherence of the system the resulting state is a sum of
classical probabilities: the ontological indeterminacies of the pure
state become gnoseological uncertainties of the mixed state. In each
instance of measurement the apparatus stays in \emph{one} of the
states $\phi_{\lambda}$ with probability $|f(\lambda)|^{2}$.
\section{conclusion}
Lead by our classical macroscopic expectations we are conditioned
towards an ontology based on fields or particles. These views failed
in the microscopic world and a compromise ontology was developed
mixing particles and field properties in a complementary way.
However this last option implies an ontology difficult, or
impossible, to imagine because reality should simultaneously have
contradicting properties of particles and fields. The proposal that
reality is made by \emph{Quantum Fields} ---extended entities in
space-time build by dynamic appearance propagation and annihilation
of virtual particles--- is compatible with the astonishing features
of quantum mechanics and can be rendered intuitive. This
interpretation of quantum mechanics follows from the formalism of
the most successful theory in physics: quantum field theory.

\end{document}